# A Proposal for a Renewed Research Emphasis in Astrophysical and Celestial Dynamics

D. J. Scheeres
Department of Aerospace Engineering Sciences
The University of Colorado at Boulder
scheeres@colorado.edu
+1-720-544-1260 (voice)

Thomas S. Statler
Department of Physics and Astronomy
Ohio University
statler@ohio.edu
+1-740-593-1722

K. Terry Alfriend, Texas A&M
Phil Armitage, University of Colorado
Joe Burns, Cornell
Michael Efroimsky, United States Naval Observatory
Alan W. Harris, Space Science Institute
Sergei Kopeikin, University of Missouri – Columbia
Marc Murison, United States Naval Observatory
Phil Nicholson, Cornell
Stanton Peale, UC-Santa Barbara
P. Kenneth Seidelmann, University of Virginia
Donald K. Yeomans, Jet Propulsion Laboratory

# A Proposal for a Renewed Research Emphasis in Astrophysical and Celestial Dynamics

Summary:
Given the impressive investment by the nation in observational Astronomy and Astrophysics facilities coming on line now and in the near future, we advocate for an increased investment in applied and fundamental research on Astrophysical and Celestial Dynamics (ACD). Specifically we call for a) continued and expanded support for applied research in ACD, b) creation of support for fundamental research in ACD and its subfields, and c) the creation of a unified program to help scientists coordinate and collaborate in their research in these fields. The benefits of this proposal are threefold. First, it will enable researchers to interpret and understand the implications of newly observed phenomena that will invariably arise from new facilities and surveys. Second, research on fundamentals will foster connections between specialists, leveraging advances found in one sub-field and making them available to others. Third, a coordinated approach for applied and fundamental research in ACD will help academic institutions in the United States to produce future researchers trained and knowledgeable in essential subfields such as Mathematical Celestial Mechanics and able to continue its advancement in conjunction with the increase in observations.

Background:
Fundamental advances in our understanding of the macroscopic universe have resulted from studies of the dynamics of particles, bodies, and continuous media in interaction with each other. The forces controlling the motion of these systems arise from a diverse set of physical processes, including not only gravitation (in the Newtonian and relativistic regimes), but also hydrodynamics, electromagnetism, radiation, and the complex physics of solid materials. Our ability to comprehend and use these processes has allowed humanity to explore the solar system, to infer the internal properties of celestial bodies, and to constrain the formation and evolution of planetary systems and the cosmos. These are exactly the goals of the field of Astrophysical and Celestial Dynamics: to explain the dynamical evolution of bodies in interaction with each other, to harness these effects in order to interpret our observations of the universe, and to enable us to travel and explore in and beyond the solar system.

The term Astrophysical and Celestial Dynamics (ACD), in this white paper, encompasses numerous sibling fields, including Galactic Dynamics (the study of galaxies and galaxy systems), Stellar Dynamics (study of stellar systems), Dynamical Astronomy (study of natural motion in general), Astrodynamics (study of controlled or artificial bodies in space environments) and Mathematical Celestial Mechanics (study of motion under Newtonian gravity and General Relativity). Developments in each of these areas have led directly to widely applicable tools and fundamental insights, for example:

- Symplectic integrators, developed in the Mathematical Celestial Mechanics community, are now used nearly ubiquitously in dynamical simulations of the Solar System, galactic systems, and cosmology.

- Density wave theory, developed by Galactic Dynamicists, has matured into a powerful analytical tool for understanding energy and angular momentum transport in black hole accretion disks, protostellar and protoplanetary disks, and planetary rings, as well as spiral structure in galactic disks.
- The Yarkovsky effect, discovered in the Astrodynamics community, has, along with studies of resonances in the restricted three-body problem, fundamentally changed our understanding of the evolution of small bodies in the Solar System.
- Dynamical systems theory, motivated by Mathematical Celestial Mechanics, has revolutionized the design of space trajectories and contributed to the development of spacecraft-based astronomical observatories of all kinds, which in turn have transformed our understanding of the cosmos.

Some of the most exciting developments in Astronomy in the most recent two decades have direct connection with ACD, including the discoveries of:

- exoplanet systems;
- the population of trans-Neptunian objects;
- the link between supermassive black holes and their host galaxies; and
- the link between X-ray binaries and globular clusters.

We are entering another "new era" of Astronomy, in which new large ground-based observatories, space-based telescopes, and massive survey projects will increase the precision and information available on astronomical bodies by orders of magnitude. To capitalize on this flood of new data will require continued investment, not only in data analysis and modeling, but also in the fundamental science behind that modeling, a large part of which has its origins in Astrophysical and Celestial Dynamics.

<u>Key Questions</u>:
We identify five major science questions most likely to drive ACD efforts in the next decade:

1. *How do planets form and migrate during their early evolution?* The discovery of numerous exoplanet systems containing "hot Jupiters" has prompted investigations into migration, driven by interactions between nascent planets and protoplanetary disks. While the major processes have been identified and simulated, the important question of what halts migration remains unanswered.
2. *How do accretion disks transport angular momentum?* Accretion disks appear in many contexts, from planetary to galactic scales. Angular momentum and energy can be transported in disks through gravitationally mediated waves, fluid viscosity, magnetic fields, and combinations of these processes. Yet a true understanding of how real disks behave has remained elusive.
3. *How have the small body populations in our Solar System evolved over time?* The orbital distribution of the trans-Neptunian objects contains clues to the early migration of the outer planets, and the near-Earth object population reveals the action of resonances and the Yarkovsky effect. The large number of binary asteroids suggests material shedding by impact, tides, or gradual spin-up, and the spin state distribution may itself be a clue to the material properties of objects that may pose a hazard to Earth.

4. *How has the merging of small galaxies into large galaxies driven the buildup of galaxy disks, spheroids, and halos?* Cosmological simulations indicate that present-day large galaxies have been built up from many smaller objects. The physics of merging, including gas dynamics and star formation, is likely responsible for the growth of the well known galaxy components, and will leave dynamical signatures in galaxy halos, detectable in upcoming large-scale surveys.
5. *How have the evolution of galaxies and the growth of supermassive black holes in their centers influenced each other?* Merging of massive galaxies is thought to lead inevitably to merging of supermassive black holes, but the route to final coalescence passes through poorly understood territory, most notably the "last parsec". Dynamics in this regime will also influence the host galaxy, altering the phase space distribution of stars and possibly fueling an active nucleus.

Significant progress on these questions will require major new developments in numerical techniques, analytic theory, and targeted observations. Many of these theoretical developments will have applications across a wide range of planetary and astrophysical problems, and may not fit into the standard disciplinary categories, divided according to wavelength, size scale, or redshift.

Proposal:

We therefore advocate continued and expanded funding for research in Astrophysical and Celestial Dynamics. Our proposal has three components:

1) Continued research investment in *applications* of ACD to the analysis and interpretation of new data and new phenomena.

2) Additional investment in *fundamental research* in ACD, through the creation of specific calls for proposals.

3) The creation of a *unified program*, with participation from NASA, NSF and academic institutions, to support and encourage research within ACD.

We expect that there will naturally be increased funding available for *applied* dynamical studies into specific results from new observatories and surveys, by way of existing programs. Given the ongoing investment in major observational resources which will generate massive datasets, the necessity of commensurate funding in applied theoretical research is clear and its importance should be well understood historically.

Additional support for *fundamental* theoretical research in ACD will help develop a more cohesive approach to this field within the United States Astronomical and Astrophysical community. Such an investment would pay off both in the near term and the longer term, nurturing an interdisciplinary community in the US that can take full advantage of the myriad individual efforts on applied problems currently being solved in the broader community.

The encouragement of research in fundamental ACD questions, paired with a wealth of new observations, can significantly leverage the investments made in observatories and surveys. Significant advances in understanding have historically emerged from the interdisciplinary sibling fields within ACD, as we demonstrated above. Many new problems in interpretation of observations can be addressed by appealing to the broad ACD community. For example, exoplanet detection has given rise to many important technical and theoretical problems, elements of which have been addressed in studies of space debris, orbit determination in Hamiltonian systems, resonance and chaos, and coupled rotational and orbital dynamics of solid bodies. However, researchers in these latter fields have no specific program that supports fundamental research, hindering the generalization of their work. Funding research on fundamental ACD problems can enable the United States to redevelop a unified academic community that can produce researchers who can compete with European and Asian institutions in producing top researchers in Mathematical Celestial Mechanics.

Finally, we propose a *unified program* devoted to ACD and dedicated to studying the dynamics of motion across all temporal and spatial scales. This research program will encourage interdisciplinary collaboration across the disparate ACD sub-communities.

First, we request continued and increased funding for applied research within each of the subfields of ACD, through traditional programs supported by NSF and NASA. In addition, we call for a new program to support fundamental research in this field. An added investment of $500K - $750K per year (perhaps 10-15 grants per year) would create an energetic response and lead to robust growth and development of new methods of analysis, simulation and cross-fertilization between the subfields of ACD. Assuming 3-year grants, the total program size once in a steady-state would be on the order of $1.5-2.2M.

The goals for this program would be:

I. Foster an interdisciplinary community that grapples with dynamics across all temporal and spatial scales, charged with applying their understanding to problems related to the motion and interaction of natural and artificial bodies.

II. Maintain NASA and NSF's traditional support of applied research in ACD through such programs as Astrophysical Theory, Planetary Geology & Geophysics, Outer Planets Research, Astrobiology, and others, and provide it with a clearer plan for its future funding.

III. Institute a funding pathway for migration of research between the different ACD sub-communities, including the development of Astrodynamics-related issues of benefit to Astronomy and Astrophysics, such as space situational awareness, mission design, trajectory navigation, and the exploitation of natural dynamics for space missions and observatories.

IV. Foster a new focus on applied problems in the Mathematical Celestial Mechanics community, tapping into the significant progress that has been made in recent years in the fields of periodic orbits, fundamental properties of solutions, and symplectic topology.

Outcomes:

A successful implementation of this program will have a number of important outcomes, including:

- Maintenance and growth of an interconnected research and academic community in the broad fields of Astrophysical and Celestial Dynamics that can serve the larger community of Astronomy and Astrophysics.
- Development of a theoretical background and foundation for tying together the disparate advances across the spectrum of research in Astronomy and Astrophysics into a more unified understanding of how the universe evolves from the largest to the smallest scales.
- Creation of healthy academic communities which can produce world-class researchers in the field of ACD, enabling the United States to be the preeminent contributor to theoretical advancements in Astronomy and Astrophysics.
- Discovery of new physical processes and their rigorous physical understanding, which could have application to many other areas of science and technology.